\documentstyle[aps,pre,epsfig,floats]{revtex}

\newcommand{\centps}[2]{
	\begin{center}
		\epsfig{file=#1,height=#2mm}
	\end{center}
}

\begin{document}

\twocolumn[\hsize\textwidth\columnwidth\hsize\csname@twocolumnfalse%
\endcsname

\draft

\title{\bf Jamming transition in a homogeneous one-dimensional system:
	the Bus Route Model}
\author{O.J.~O'Loan, M.R.~Evans and M.E.~Cates\\}
\address{Department of Physics and Astronomy, University of Edinburgh,
      Mayfield Road, Edinburgh EH9 3JZ, U.K.\\}
\maketitle

%###############################################################################
% Abstract
%###############################################################################

\begin{abstract}
We present a driven diffusive model which we call the Bus Route Model. The
model is defined on a one-dimensional lattice, with each lattice site having
two binary variables, one of which is conserved (``buses'') and one of which is
non-conserved (``passengers''). The buses are driven in a preferred direction
and are slowed down by the presence of passengers who arrive with rate
$\lambda$.  We study the model by simulation, heuristic argument and a
mean-field theory. All these approaches provide strong evidence of a transition
between an inhomogeneous ``jammed'' phase (where the buses bunch together) and
a homogeneous phase as the bus density is increased.  However, we argue that a
strict phase transition is present only in the limit $\lambda \to 0$. For small
$\lambda$, we argue that the transition is replaced by an abrupt crossover
which is exponentially sharp in $1/\lambda$. We also study the coarsening of
gaps between buses in the jammed regime. An alternative interpretation of the
model is given in which the spaces between ``buses'' and the buses themselves
are interchanged. This describes a system of particles whose mobility decreases
the longer they have been stationary and could provide a model for, say, the
flow of a gelling or sticky material along a pipe.
\end{abstract}


\vspace{1cm}
]

\section{Introduction and Model}		\label{sec:Introduction}

Driven diffusive systems \cite{SZ} have recently attracted much attention in the
field of non-equilibrium statistical mechanics from a fundamental viewpoint, as
well as in the context of traffic modelling \cite{Nagel}, interface
growth \cite{HHZ} and other applications \cite{LR}.

One particularly interesting feature is the possibility of phase ordering and
phase transitions in one dimensional ($1d$) systems. To appreciate the
significance, one should recall that in $1d$ equilibrium models, ordering only
occurs either in the limit of zero temperature ({\it e.g.} kinetic Ising models
or deterministic Ginsburg-Landau equation) or in mean-field-like models.
However, in non-equilibrium systems it has been demonstrated that ordering may
occur in models with fully stochastic, local dynamics \cite{Gacs,EFGM,GLEMSS}.

The non-equilibrium transitions found so far appear to be of three main types
(although very recently, novel phase separation phenomena have been
demonstrated in some $1d$ systems \cite{AHR,EKKM}). First there are boundary
induced transitions \cite{Krug,EFGM}. These occur on open systems with a
dynamics that conserves some quantity in the bulk, but that allows injection
and extraction of the quantity at the boundaries. A second class of transition,
describing roughening of a $1d$ interface, is connected to directed percolation
and corresponds to a driven diffusive system with non-conserved order
parameter\cite{AEHM,KW}. Finally, there are transitions induced by defect sites
\cite{JL,Schutz} or particles \cite{DJLS,Mallick,Derrida,TZ} or the presence of
disorder \cite{Evans,KF,BK}. In this class of systems, the presence of the
defect causes a macroscopic region of high density to form. Analogies with Bose
Einstein condensation \cite{Evans} and phase coexistence of a gas
\cite{Derrida} have been made. An even simpler way to view the phenomenon is as
a jamming transition; the defect causes a traffic jam to form behind it. In
this context, however, ``jamming'' may be a somewhat misleading term, since in
the models just described the inhomogeneous ``jammed'' phase arises at {\em
low} density. The transition is in fact between this phase and a higher density
``congested'' phase which is uniform, but which has a mean particle velocity
{\em lower} than that of the jammed phase. Indeed, it appears that a minimum
velocity principle applies \cite{Evans}, so that the stable phase is always
the slowest available at a given mean density.

In this work we address the question of whether similar ``jamming'' transitions
can occur in $1d$ homogeneous systems, {\it i.e.\/} systems with periodic
boundary conditions and without disorder or defects. We introduce a model that
exhibits a jamming transition in a certain limit (to be specified below). For
the moment, it is useful to describe the model in terms of a commonly
experienced and universally irritating situation. Consider buses moving between
bus-stops along a bus route. Clearly, the ideal situation is that the buses are
evenly distributed along the route so that each bus picks up roughly the same
number of passengers. However, owing to some fluctuation, it may happen that a
bus is delayed and the gap to the bus in front of it becomes large. Then, the
time elapsed since the bus-stops in front of the delayed bus have been visited
by the previous bus is larger than usual and consequently more passengers will
be waiting at these bus-stops. Therefore the bus becomes delayed even further.
At the same time, the buses behind catch up with the delayed bus and pick up
only very few passengers since the delayed bus takes them all. Hence a ``jam''
of buses forms. Inspired by this scenario we shall formulate a model below, to
be referred to as the Bus Route Model (BRM).

We defer the mathematical definition of the BRM until after we have
discussed the general context. Already, from the simple picture discussed in
the previous paragraph, we can identify a conserved variable (the buses) and a
non-conserved variable (the passengers). The passengers are non-conserved since
they arrive at the bus-stops from outside the system (bus route). The
conserved buses are driven in a preferred direction. We may usefully think of
the non-conserved variable coupling to the conserved variable and mediating the
jamming transition.

Associated with any ordering dynamics is the phenomenon of coarsening
\cite{Bray} where the typical domain length of the ordered phase grows
indefinitely with time. Indeed, coarsening has been studied in ballistic
aggregation models \cite{BKR} and disordered driven diffusive systems where
jamming occurs \cite{KF,Nagatani}. Again, a contrast can be made with $1d$
equilibrium models where only zero temperature models or mean-field-like models
coarsen. In the present model, it is the gaps between the jams that coarsen as
the jams aggregate; we study this phenomenon in the present work.

In a finite system, the coarsening eventually results in one large jam with a
single gap in front of it. Recalling that the model system is homogeneous and
that no bus is preferred over any other, we see that we have a spontaneous
symmetry breaking where the symmetry between buses is broken through one bus
being selected to head the jam. Symmetry breaking transitions have been
previously been found in $1d$ open systems with a non-conserved variable at the
boundaries \cite{EFGM,GLEMSS} and in a class of growth models
\cite{AEHM}. The present model provides an example in a homogeneous system with
a conserved variable. Related symmetry breaking has also been noted in some
models in a class of ``Backgammon'' or ``balls in boxes'' models which are
effectively simple generalisations of Bose systems \cite{Ritort,BBJ,DGC}.
However, in these models the dynamics is inherently equilibrium and
mean-field-like whereas the BRM has local dynamics that does not satisfy
detailed balance. We shall elucidate the connection between the two classes of
models by showing that a mean-field approximation to the BRM results in a model
that may be solved analytically. The steady state of this soluble mean-field
model falls into the class of generalised Bose systems.

We now formally define the BRM. The model is defined on a
$1d$ lattice with periodic boundary conditions. Each lattice site is
labelled by a number $i$ running from $1$ to $L$. Site $i$ has two binary
variables $\tau_i$ and $\phi_i$ associated with it. These variables can be
described in the following terms:
\begin{itemize}
	\item If site $i$ is occupied by a bus then $\tau_i = 1$; 
	otherwise $\tau_i = 0$. 
	\item If site $i$ has passengers on it then $\phi_i = 1$;
	otherwise $\phi_i = 0$.
\end{itemize}
Each site can be thought of as a bus-stop on a bus route. A site cannot have
both $\tau_i = \phi_i = 1$ ({\it i.e.\/} it cannot have a bus {\em and}
passengers).

There are $M$ buses and $L$ sites in the system and the bus density 
\begin{equation}
	\rho = M/L
	\label{eqn:rho-def} 
\end{equation}
is a conserved quantity. However, the total number of sites with passengers is
{\em not} conserved. The update rules for the system are as follows:
\begin{enumerate}
	\item Pick a site $i$ at random.
	\item If $\tau_i = 0$ and $\phi_i = 0$ then $\phi_i \to 1$ with probability
	$\lambda$.
	\item If $\tau_i = 1$ and $\tau_{i+1} = 0$, define a hopping rate $\mu$ 
	as follows: 
		\begin{itemize} 
			\item $\mu = \alpha \;$ if $\; \phi_{i+1} = 0$
			\item $\mu = \beta \;$ if $\; \phi_{i+1} = 1$  
		\end{itemize} 
	and update $\tau_i \to 0$, $\tau_{i+1} \to 1$ and $\phi_{i+1} \to 0$ 
	with probability $\mu$.
\end{enumerate}

\begin{figure}			
	\centps{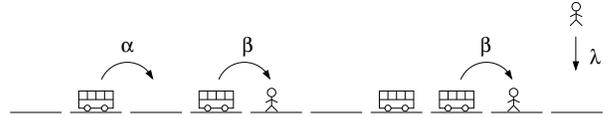}{15}
	\caption{Schematic illustration of the BRM.}
	\label{fig:Schematic}
\end{figure}

Thus, $\alpha$ is the hopping rate of a bus onto a site with no passengers and
$\beta$ is the hopping rate onto a site with passengers. The probability that a
passenger arrives at an empty site is $\lambda$. When a bus hops onto a site
with passengers, it removes the passengers. While we have taken the passenger
variable $\phi_i$ to be binary, this does not forbid the presence of more than
one passenger at a site; we merely require that the extra passengers have no
further effect on the dynamics. We generally take $\beta < \alpha$, reflecting
the fact that buses are slowed down by having to pick up passengers. We may set
$\alpha$ equal to $1$ without loss of generality and from now on we consider
only this case. We note that the dynamics is local and does not satisfy
detailed balance. The model is illustrated schematically in Figure
\ref{fig:Schematic}.

Although the language of ``buses'' and ``passengers'' provides an appealing
mental picture, the model is intended to be simple rather than realistic. For
example, the need for buses to stop to allow passengers to disembark is
ignored. Note however, that an ability for buses to overtake each other would
have almost no effect. This is because, in a jammed situation of the type
discussed below, the interchange of a fast-moving bus with a slower-moving one
in front also interchanges their velocities.

The main purpose of this paper is to provide evidence that the model defined
above undergoes a jamming transition. However, we argue that a strict
transition ({\it i.e.\/} a singularity in some measured quantity) only occurs as
$\lambda \to 0$ and in the thermodynamic limit. Here we define the
thermodynamic limit as
\begin{equation}
	M, \; L \; \to \; \infty \;\;\; \mbox{with} \;\;
	\rho \;\; \mbox{held fixed.} 
	\label{eqn:ThermoLimit}
\end{equation}

Our evidence is both numerical and analytical. In Section \ref{sec:Results} we
present Monte Carlo simulations which provide evidence for the above
picture. In the following sections we provide various analyses to support the
picture. In particular, in Section \ref{sec:TwoBody} we present a simple two
particle approximation which describes the stability of jams. In Section
\ref{sec:MFM} we define a mean-field approximation. Within this approximation
we can solve analytically for the steady state and find it is similar to a
generalised ideal Bose gas or Backgammon model. This steady state can be
analysed and exhibits a phase transition (taking the form of a condensation
transition) only as $\lambda \to 0$ in the thermodynamic limit. We show that
the mean-field steady state agrees quantitatively with Monte Carlo simulations
of the BRM, suggesting that for the BRM also there is no strict transition for
nonzero $\lambda$. We also study numerically the behaviour as $\lambda \to 0$
in the mean-field approximation.

In Section \ref{sec:Coarsening} we study the approach to the steady state in
the jammed phase where we observe coarsening of bus clusters. We argue that on
an infinite system the size of the large gaps between clusters should
eventually grow as $t^{1/2}$. We study finite systems numerically. In Section
\ref{sec:Holes} we discuss an interpretation of the BRM wherein we consider 
the vacant sites {\em between} buses to be the moving entities in the
system. Viewed this way, the non-conserved variable now describes an internal
degree of freedom of the moving entities themselves, rather than of the sites
they visit. We discuss possible physical interpretations of this dynamics,
including a model of clogging. In Section \ref{sec:RealBuses}, we consider the
relevance of the BRM to real bus routes and in Section
\ref{sec:Discussion}, we conclude with a discussion of the main points of our
work.

\section{Simulation Results and Heuristic Arguments} \label{sec:Results}

The model defined above captures an important feature of the bus route problem
described in the introduction, namely that once a gap between buses becomes
large through some fluctuation, the tendency is for the gap to become still
larger: since buses move more quickly in areas with few passengers, a bus which
is following closely behind another will tend to move faster than one which is
a long way behind the bus directly in front. This is simply because the closer
a bus is to the one directly ahead of it, the less time passengers will have
had to arrive. If this tendency for large gaps to grow were to prevail, the
result on a finite system would be a single jam of buses and one large gap. We
first argue that in the limit of $\lambda \to 0$ but $\lambda L \to \infty$
this scenario can only hold at low enough density of buses, and that a phase
transition to a homogeneous phase will result as the density is raised.

First consider a system comprising a single large jam. In order for this to be
a stable object, the velocity (defined as the average rate of hopping forward)
of the leading bus must be equal to the velocity of any bus inside the jam. Now
if $\lambda L \to \infty$, the probability that the site immediately in front
of the leading bus has a passenger on it tends to one. (This is because the
rate of passenger arrival, multiplied by the time delay between the final bus
of the jam and the leading bus of the jam crossing the same site, is of order
$\lambda L$). Therefore the velocity of the leading bus will approach
$\beta$. On the other hand, if $\lambda \to 0$ the probability that a site
within the jam has passengers tends to zero since these gaps are of finite
length. Therefore, since the velocity of a bus in the jam will only be limited
by the presence of neighbouring buses, this velocity will be $1-
\rho_{\rm jam}$ where $\rho_{\rm jam}$ is the density of buses {\it within} the
jam\footnote{This follows from the fact that, when $\lambda \to 0$ with
$\lambda L \to \infty$, the situation is equivalent to a model of hopping
particles with a single slow ``defect'' particle \cite{Evans}.}. Equating the
two velocities yields $\rho_{\rm jam}= 1-\beta$. However, the jam must clearly
have density greater than $\rho$, the overall density of the system. Thus for
jamming to occur we require
\begin{equation}
	\rho < \rho_c \;\;\; \mbox{where} \;\;\; 
	\rho_c = \rho_{\rm jam} = 1 - \beta.
\label{naive}
\end{equation}
For densities above this critical value, the system will be in a congested
phase where gaps between buses are uniform.

An equivalent way of obtaining (\ref{naive}) is to compare the velocity in the
jammed phase, $\beta$, and the velocity in the homogeneous phase, $1-
\rho$. The phase with the lowest velocity is chosen by the system. This
procedure is analogous to the thermodynamic procedure of choosing the phase
with the lowest chemical potential. Though unproven for the present problem, in
the case of disordered exclusion processes the analogy has been shown to be
exact \cite{Evans}.

\begin{figure}
	\centps{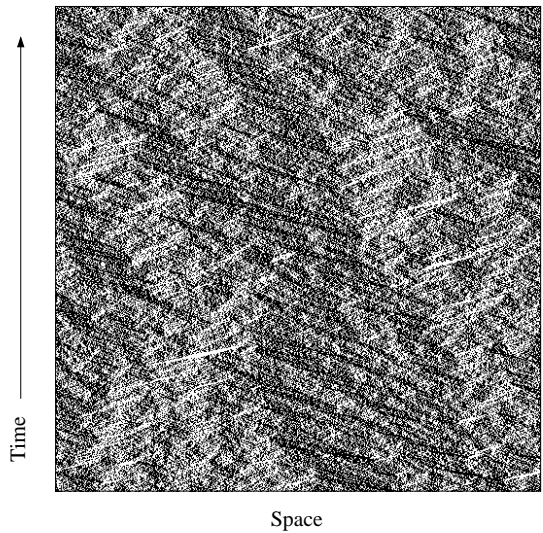}{70} 
	\caption{Space-time plot of bus positions for $\lambda = 0.02$, $\rho =
	0.55$, $\beta = 0.5$ and $L=500$. There are 10 time-steps between each
	snapshot on the time axis. Initially the buses are positioned randomly and
	there are no passengers.}
	\label{fig:ST-0.02-nojam}
\end{figure}

\begin{figure}
	\centps{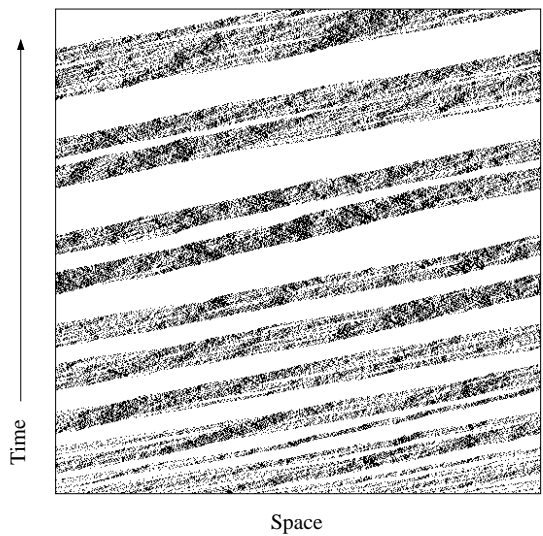}{70} 
	\caption{Space-time plot of bus positions for the same parameters as in
	Figure \ref{fig:ST-0.02-nojam} except that here, $\rho = 0.2$.}
	\label{fig:ST-0.02-jam}
\end{figure}

We now present simulations which, for small $\lambda$, qualitatively support
the above picture of a phase transition. Figure \ref{fig:ST-0.02-nojam} shows a
space-time plot of the buses in system at $\rho = 0.55$ for small $\lambda$
($\lambda = 0.02$). The buses are distributed fairly homogeneously throughout
space. There are no very large gaps present in the system. Figure
\ref{fig:ST-0.02-jam} shows a space-time plot for $\rho = 0.2$ which is less
than $\rho_c$ of (\ref{naive}). In this case, starting from a random
configuration of buses, large gaps quickly open up and small clusters or
``jams'' of buses are readily seen. Gradually, these small jams coarsen until,
finally, the system comprises a single large jam. There is one large inter-bus
gap in front of the jam whereas the jam itself contains many small gaps. The
behaviour can be thought of as phase separation into regions of nonzero and
zero density.

\begin{figure}
	\centps{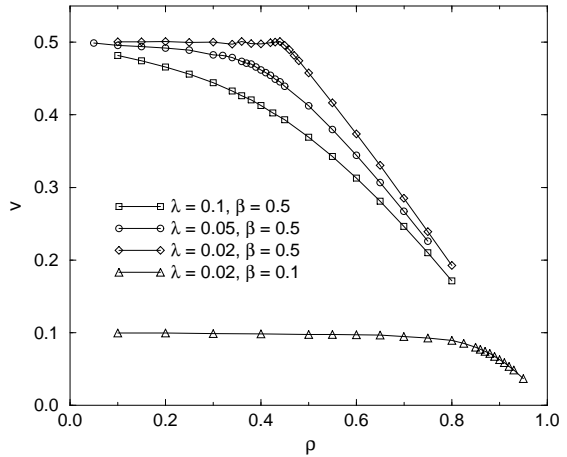}{70}
	\caption{Velocity as a function of density for various values of $\lambda$
	and $\beta$. The simulations were performed with $L=10000$. The lines are
	shown to guide the eye.}
	\label{fig:SD-1}
\end{figure}

In order to investigate the effect of varying $\lambda$, we plot in Figure
\ref{fig:SD-1} the steady state average velocity $v$ as a function of the
density. A system size $L = 10000$ was chosen since using a bigger system did
not appreciably affect the results. Let us first consider the results for
$\beta = 0.5$. For $\lambda = 0.1$, the velocity increases smoothly as the
density is decreased and appears to approach $\beta$ for small density. There
is no sign of a phase transition. For $\lambda = 0.05$ a similar picture holds
although now the curve $v(\rho)$ is more concave.

A strikingly different picture is obtained for $\lambda = 0.02$. There is an
apparent discontinuity in the derivative of $v(\rho)$ at some value of the
density $\rho^* \simeq 0.45$. Below $\rho^*$, the data is consistent with $v =
\beta$ whereas above $\rho^*$, $v$ decreases almost linearly with increasing 
density. This behaviour is consistent with the simple picture of a jamming
transition discussed above. It is also very similar to the velocity-density
relationship in an exactly solvable model with a single slow particle
\cite{Evans,KF} where a jamming transition occurs. The graph therefore suggests
that a phase transition may occur for small $\lambda$. We argue, however, that
the phase transition occurs only in the limit $\lambda \to 0$ and that this
limit has to be taken in an appropriate way. In order to quantify this we now
analyse a simple two-particle approximation to the full system.

\subsection{Two-Particle Approximation}	\label{sec:TwoBody}

\begin{figure}
	\centps{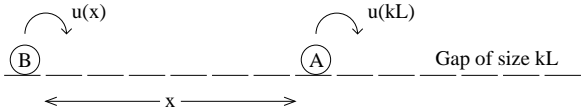}{14}
	\caption{ The two buses at the head of a jam in a system of size L.}
	\label{fig:GapHead}
\end{figure}

Consider a system containing $M$ buses. Let us assume that there is a jam in
the system ({\it i.e.\/} a gap with size ${\mathcal O} (L)$). If there is a
jamming transition then such a gap should become stable in the thermodynamic
limit; the bus at the head of the jam should not be able to escape into the
large gap. Let us consider the two buses $A$ and $B$ at the head of such a jam
as shown in Figure \ref{fig:GapHead}.
The gap in front of $A$ has size $kL$ where $k$ is independent of $L$. The gap
in front of $B$ has size $x$. We now assume that we can write the hop rate of
either bus as a function only of the gap size in front of that bus. To do this
we write $u(x) = f(x) + \beta (1-f(x))$ where $f(x)$ is the probability that
there are no passengers on the first site of a gap of size $x$. To estimate
$f(x)$ we assume that bus $A$ is a random walker hopping with rate $v$, where
$v$ is the average velocity of the system. Then the average time since $A$ left
the site in front of $B$ is $x/v$. Since the rate of arrival of passenger is
$\lambda$ we then have $f(x) = \exp(-\lambda x/v)$ and
\begin{eqnarray}
u(x) &=& \beta + (1-\beta)\exp(-\lambda x/v)\;\;\;
\mbox{for}\;\;\;x >0 \nonumber \\
u(0)& =& 0.
\label{eqn:HopProb}
\end{eqnarray}
The use of expression (\ref{eqn:HopProb}) for the hopping rates is in the
spirit of a mean-field approximation. We have found by simulation that the
approximation is a good one for $\beta$ larger than about $0.2$, but for small
$\beta$ it breaks down. In particular, for small $\lambda$ (when $\beta$ is
small) we find that $u(x)$ decays to $\beta$ much more rapidly than
(\ref{eqn:HopProb}) predicts. We believe that the reason for this is the
failure of (\ref{eqn:HopProb}) to take into account time correlations in the
hopping of buses -- when a bus is updated and fails to hop into an unoccupied
site, the next time it is updated it should hop with probability $\beta$ and
{\em not} $u(x)$ (because for a bus to fail to hop, the site ahead {\em must}
contain passengers\footnote{This is no longer true for $\alpha \neq 1$, but
time correlations will still be present.}). Clearly, the effects of neglecting
these time correlations will be largest for small $\beta$. We shall return to
(\ref{eqn:HopProb}) in Section \ref{sec:MFM} where we carry out a conventional
mean field theory for the many particle problem. For our present purposes, we
replace $v$ by $\beta$ in (\ref{eqn:HopProb}) since we are interested in a
jammed situation.

Using the mean field hopping rate $u(x)$ we may write a Langevin equation for
the dynamics of the gap size
\begin{equation}
	\dot{x} = u(kL) - u(x) + \eta(t)
\end{equation}
where $\eta(t)$ is a noise term (say white noise of unit variance\footnote{The
variance of the noise should strictly depend on $\beta$ but, since we are
primarily interested in the effect of (small) $\lambda$ on the dynamics of the
gap, we ignore this dependence.}). This can be written in the form
\begin{equation}
	\dot{x} = -\frac{d}{dx} \Phi(x) + \eta(t)
\end{equation}
where
\begin{equation}
\Phi(x) = -(1 -\beta) \left[
x {\rm e}^{-\lambda k L/\beta} + 
	\frac{\beta}{\lambda} {\rm e}^{-\lambda x/\beta}
\right].
\end{equation}
The gap size $x$ has the dynamics of a particle undergoing diffusion in a
potential $\Phi(x)$ for $x>0$. There is a reflecting boundary at $x=0$ and the
potential has a maximum at $ x^* = kL$. Therefore for $x<x^*$ the particle is
trapped in a well ($ 0 < x < x^*$) and bound to the reflecting boundary, {\it
i.e.\/} $A$ is bound to the head of the jam. If $x > x^*$, the particle has
escaped from the well and $A$ is no longer at the head of the jam. We denote
the time to escape from the well by $\tau$ which, to a first
approximation \cite{Kramers}, is given by
\begin{equation}
	\tau \sim \exp \left( \Phi(x^*) - \Phi(0) \right).
	\label{tau1}
\end{equation}

In the thermodynamic limit $L \to \infty$, we have
\begin{equation}
	\tau \sim \exp \left[ 
		\frac{\beta(1-\beta)}{\lambda} \right]
	\label{tau2}
\end{equation}
which is finite for $\lambda > 0$, showing that particle $A$ will leave the
head of the jam in a finite time. This implies that a jam is not a stable
object and will eventually break up. However, when $\lambda \to 0$, particle
$A$ becomes bound to particle $B$ since $\tau$ diverges. Therefore, in this
limit, the jam is a stable object. This simple analysis suggests that there is
a phase transition (between a jammed and a homogeneous phase as the density is
raised) only in the limit $\lambda \to 0$. When $\lambda$ is small but nonzero,
$\tau$ is exponentially large in $1/\lambda$ and it can appear that a jam is
stable when in fact it has a finite lifetime.

It is clear that the limits $\lambda \to 0$ and $L \to \infty$ do not commute
since as $\lambda \to 0$ on a finite system one recovers a model where all
hopping rates are the same. Such a model of hopping particles with hard core
exclusion (known as an asymmetric exclusion process) has been well studied and
exhibits no phase transition with periodic boundary conditions \cite{SZ}. In
practice, one could take the limit $L \to \infty$ and then $\lambda \to 0$ by
choosing
\begin{equation}
\lambda \sim L^{-\gamma}\;\;\;\mbox{with}\;\;\; \gamma < 1
\end{equation}
and taking the thermodynamic limit $L\to \infty$ whereupon the escape time
diverges as
\begin{equation}
	\tau \sim \exp(aL^{\gamma})
\end{equation}
and the jam is a stable object.

\begin{figure}
	\centps{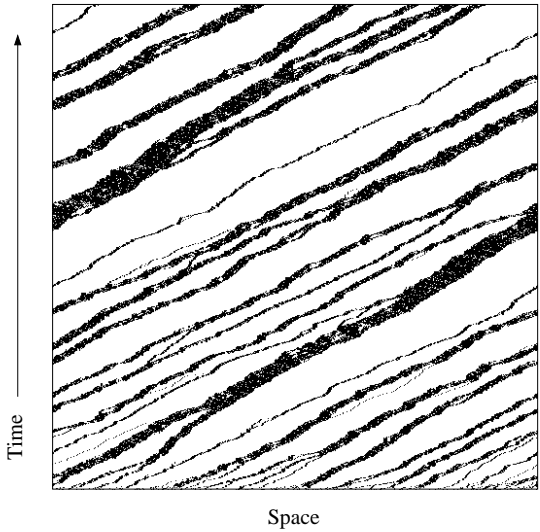}{70} 
	\caption{Space-time plot of bus positions for $\lambda = 0.02$, $\rho =
	0.2$, $\beta = 0.1$ and $L=500$. There are 20 time-steps between each
	snapshot on the time axis. Initially the buses are positioned randomly and
	there are no passengers.}
	\label{fig:ST-0.1-0.02}
\end{figure}
In order to test this picture against simulations, we plot in Figure
\ref{fig:ST-0.1-0.02} a space-time plot for $\beta = 0.1$ and $\lambda = 0.02$
for a low density. At first glance, it appears that jamming just like that seen
for $\beta = 0.5$ is taking place. However, closer inspection of the individual
jams shows that they can develop large gaps and, in some cases, divide into
smaller jams. The steady state for this system is therefore not characterised
by a single large jam (as appeared to be the case from simulations for $\beta =
0.5$), but comprises a number of jams of varying sizes.

\begin{figure}
	\centps{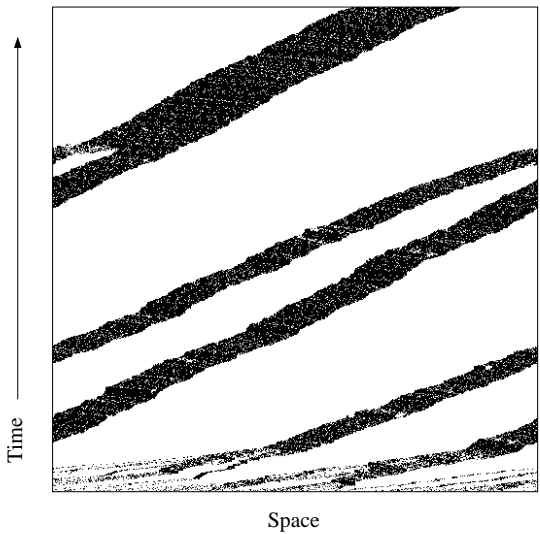}{70}
	\caption{Space-time plot of bus positions for the same parameters as in
	Figure \ref{fig:ST-0.1-0.02} except that here, $\lambda = 0.002$.}
	\label{fig:ST-0.1-0.002}
\end{figure}
Figure \ref{fig:ST-0.1-0.002} shows a space-time plot for the same value of
$\beta$ ($0.1$) but for $\lambda = 0.002$, a factor of 10 smaller. It appears
that the jamming has been restored -- the individual jams are not seen to break
up. It is also interesting to note that splitting of jams is not observed for
$\beta = 0.5$ (Figure \ref{fig:ST-0.02-jam}) so the splitting in Figure
\ref{fig:ST-0.1-0.02} is a direct consequence of having a smaller value of
$\beta$. Figure \ref{fig:SD-1} also shows $v(\rho)$ for $\lambda = 0.02$ and
$\beta = 0.1$. Unlike the case $\beta = 0.5$, $v(\rho)$ is a smooth function
with apparently no discontinuity in its derivative. This is consistent with the
observation that the jammed state is unstable on the time scale of the
simulation (see Figure \ref{fig:ST-0.1-0.02}).

We therefore conclude from our two-particle argument that although the
simulation data is indicative of the presence of a phase transition, it cannot
be taken as firm evidence that a strict transition occurs for any nonzero
$\lambda$. However, for small $\lambda$ something closely approaching a
transition is seen.

%###############################################################################
% Mean Field Theory
%###############################################################################

\section{Mean-Field Model}		\label{sec:MFM}

In this section we discuss a mean-field theory which we believe describes the
behaviour of the BRM very well. Motivated by the mean-field probability
(\ref{eqn:HopProb}) of hopping into a gap of a given size, derived in Section
\ref{sec:TwoBody}, we approximate the BRM by a model which is exactly solvable
in the steady state. We term this new model the {\em mean-field model}
(MFM). We analyse this MFM and show that it compares quantitatively well with
the BRM and, importantly, that it does not have a phase transition for $\lambda
> 0$. However, we find that there is a transition in the MFM in the limit
$\lambda \to 0$ so long as the thermodynamic limit is taken first. This
supports our view that the same applies in the BRM itself.

We define the MFM to be a hopping particle model (generalised asymmetric
exclusion process \cite{SZ}) with the hopping probability of the $i^{\rm th}$
particle being $u(x_i)$, a function of $x_i$, the size of the gap in front of
that particle. As in the BRM, we consider $M$ particles and $L$ sites in total,
and periodic boundary conditions apply.  The hopping probabilities $u(x)$ are
given by the mean-field expression (\ref{eqn:HopProb}). Note that $u(x)$
depends on $v$ which is the steady state mean velocity of the system, $\langle
u \rangle$. It is useful to think of $v$ in (\ref{eqn:HopProb}) as an
adjustable parameter in the MFM to be determined self-consistently; {\it i.e.\/}
$v$ is chosen (for fixed $\rho$, $\beta$ and $\lambda$) so that $\langle u
\rangle = v$.

While $u(x)$ is given by (\ref{eqn:HopProb}) for the MFM, it is convenient to
first consider a more general $u(x)$, with the only constraints being $u(0) =
0$ (exclusion) and $u(x) \to \beta^+$ as $x \to \infty$.  When one considers
the dynamics of the gaps between particles, it is evident that such a model is
an example of what is known in the mathematical literature as a zero-range
process \cite{Spitzer,Andjel} which is exactly soluble (in the
steady-state). We present the solution in detail in Appendix
\ref{sec:ExactSoln} and quote the result here. Each configuration can be
uniquely identified by a set of gap sizes $\{ x_i \} =
\{ x_1, x_2, \ldots, x_M \}$. The steady-state probability of a configuration
$\{ x_i \}$ is
\begin{equation}
	P( \{ x_i \} ) = \frac{1}{Z(L,M)} q(x_1) q(x_2) \cdots q(x_M)
	\label{eqn:ExactSoln}
\end{equation}
where
\begin{equation}
	q(x) = \prod_{y=1}^{x} \frac{1}{u(y)}
	\label{eqn:Define_q}
\end{equation}
and
\begin{equation}
	Z(L,M) = \sum_{x_1, x_2, \ldots, x_M} 
				q(x_1) \cdots q(x_M) 
				\delta_{x_1 + \ldots + x_M, L - M}
	\label{eqn:Z1}
\end{equation}
$Z(L,M)$ can be viewed as the partition function of a generalised
non-interacting system of Bose particles. We follow Bialas, Burda \& Johnston
\cite{BBJ} in the consequent analysis, starting from the expression for
$Z(L,M)$. We first discuss the behaviour for general $u(x)$ in the
thermodynamic limit and state the conditions which $u(x)$ must satisfy in order
for a phase transition to occur. Turning to the specific case of the MFM, we
show that there is a phase transition in the limit $\lambda \to 0$. We
describe a method for analysing finite systems and we compare the MFM with
simulation results for the BRM. Finally, we discuss the behaviour for small,
but nonzero, $\lambda$.

\subsection{Mean-Field Model in the Thermodynamic Limit}	
	\label{sec:MFM-analysis}

Using the integral representation of the Kronecker delta
\[
	\delta_{m,n} = \oint \frac{ds}{2 \pi {\rm i}} \frac{s^m}{s^{n+1}}
\]
we write $Z(L,M)$, defined in (\ref{eqn:Z1}), as
\begin{equation}
	Z(\rho,M) = \oint \frac{ds}{2 \pi {\rm i} s} 
		\left[ \frac{F(s)}{s^{\frac{1}{\rho}-1}} \right]^M 
	\label{eqn:Z2}
\end{equation}
where $\rho = M/L$ and we have defined the generating function
\begin{equation}
	F(s) = \sum_{x=0}^{\infty} q(x) s^x.
	\label{eqn:Fdef}
\end{equation}

The integral (\ref{eqn:Z2}) is calculated in the thermodynamic limit (defined
in (\ref{eqn:ThermoLimit})) using the saddle point method. It can be shown that
the saddle point, where $s = z$, is given through the expression
\begin{equation}
	\frac{1}{\rho}-1 = z g(z)
	\label{eqn:Saddle}
\end{equation}
where 
\begin{equation}
	g(z) = \frac{F'(z)}{F(z)}. 
\end{equation}
We may identify $z$ as the fugacity. Each value of the fugacity gives a
particular value of the density. The partition function becomes
\begin{equation}
	Z(\rho,M) \sim \exp \left\{ M \log F(z) - 
		(L-M) \log(z) \right\}
	\label{eqn:SaddleZ}
\end{equation}
with $z$ for a particular $\rho$ determined by solving (\ref{eqn:Saddle}).

To obtain quantitative results from this solution, it is useful to obtain an
expression for $p(x)$, the steady state probability that a given particle has a
gap of size $x$ in front of it. For any $L$ and $M$, $p(x)$ is given by
\begin{eqnarray}
	p(x) &=& \frac{q(x)}{Z(L,M)}
		\sum_{x_2, \ldots , x_M} q(x_2) \cdots q(x_M) 
		\delta_{x_2 + \ldots + x_M, L - M - x}
		\nonumber \\
	&=& q(x) \frac{Z(L-x-1,M-1)}{Z(L,M)}.
	\label{eqn:Pcalc}
\end{eqnarray} 
This can be determined for a finite gap size $x$ in the thermodynamic limit
using the saddle-point expression for $Z(L,M)$ given in (\ref{eqn:SaddleZ}). We
obtain
\begin{equation}
	p(x) = \frac{q(x) z^x}{F(z)}.
	\label{eqn:Induced}
\end{equation}
This expression is in terms of $z$ but $\rho$ may be found using 
(\ref{eqn:Saddle}).

The mean particle velocity $v$ in the steady state is
\begin{equation}
	v = \langle u \rangle = \sum_{x=1}^{\infty} u(x) p(x).
	\label{eqn:u-average}
\end{equation}
Substituting (\ref{eqn:Induced}) into (\ref{eqn:u-average}) gives the result
that $v = z$. This is a relationship which has been found before in this kind
of system \cite{Evans}. We show below that $z$ is constrained to be no greater
than $\beta$ and hence, there is an upper limit on the velocity. This
constraint on $z$ combined with (\ref{eqn:Induced}) implies that in the
thermodynamic limit, $p(x)$ is a monotonically decreasing function of $x$ and
behaves for large $x$ as $p(x) \propto (z/\beta)^x$.

We now examine the criteria that $u(x)$ must satisfy for a phase transition to
occur. This entails analysis of the generating function $F(z)$ defined by
(\ref{eqn:Fdef}). $F(z)$ converges for $z < \beta$ and diverges for $z >
\beta$ (since we have $u(x) \to \beta^+$ as $x \to \infty$). Now, 
$g(z=0) = 0$ which corresponds to $\rho = 1$. Observe that $g(z)$ is a
monotonically increasing function for $0 \le z \le \beta$ which means that
$\rho$ decreases monotonically from $1$ to $1/(1+\beta g(\beta))$ for $z$ in
this range. It is clear then that $g(\beta)$, if finite, will give a critical
density, $\rho_c$, via (\ref{eqn:Saddle}). Indeed, Bialas, Burda and Johnston
\cite{BBJ} show that in this case there is a transition from a high density
``congested'' phase to an inhomogeneous ``jammed'' phase for $\rho < \rho_c$
where a single gap will occupy a finite fraction of the sites in the
system. The critical density is given by
\begin{equation}
	\rho_c = \lim_{z \to \beta^-} \frac{1}{1+z g(z)}.
	\label{eqn:rhocrit}
\end{equation}
For a transition to occur at nonzero density, we must have $g(z)$ (and hence
both $F(z)$ and $F'(z)$) convergent as $z \to \beta^-$.

For convenience, we write $u(x) = \beta ( 1 + \zeta(x) )$. By considering the
asymptotic behaviour of the product $q(x)$ in the large $x$ limit, one finds
that a phase transition occurs ({\it i.e.\/} $\rho_c$ is nonzero) if and only if
$\zeta(x)$ decays to zero more slowly than $2/x$ as $x \to \infty$. For the MFM,
where $u(x)$ is given by (\ref{eqn:HopProb}), we have $\zeta(x) \sim
\exp(-\lambda x)$ and we immediately see that there is no transition
for $\lambda > 0$. We now show that, in contrast, in the limit $\lambda \to
0^+$, a transition does occur.

For the MFM, $F(z)$ is given by
\begin{equation}
	F(z) = \sum_{x=0}^{\infty} \left( \frac{z}{\beta} \right )^x
		\prod_{y=1}^{x} \frac{1}{1 + \zeta(y)} 
	\label{eqn:Fdef2}
\end{equation}
where $\zeta(x) = (1/\beta - 1) \exp ( -\lambda x / v )$. To show that a
transition occurs in the limit $\lambda \to 0^+$, we must show that
\begin{equation}
	\lim_{z \to \beta^-} \lim_{\lambda \to 0^+} F(z)  
	\;\;\;\; \mbox{and} \;\;\;\;
	\lim_{z \to \beta^-} \lim_{\lambda \to 0^+} F'(z)  
	\label{eqn:show1}
\end{equation}
both converge and also that
\begin{equation}
	\lim_{z \to \beta^+} \lim_{\lambda \to 0^+} F(z) =  
	\lim_{z \to \beta^+} \lim_{\lambda \to 0^+} F'(z) = \infty. 
	\label{eqn:show2}
\end{equation}
Since for $\lambda > 0$ both $F(z)$ and $F'(z)$ diverge when $z \ge \beta$, the
requirement (\ref{eqn:show2}) is satisfied trivially. Note that if the limit
$\lambda \to 0$ is taken {\em before} the thermodynamic limit, values of the
fugacity greater than $\beta$ are permitted and we recover a model where all
particles hop with probability $1$.

We now show that the expressions in (\ref{eqn:show1}) are finite and we
calculate the critical density. Clearly, $F(z)$ and $F'(z)$ both converge for
$\lambda \ge 0$ when $z < \beta$ (since the geometric part $(z/\beta)^x$
dominates the series for large $x$). Since $1 + \zeta(x) = 1/\beta$ for $\lambda
= 0$, one can easily calculate $F(z) = \sum z^x$ and $F'(z) = \sum x z^{x-1}$.
Using the saddle point equation (\ref{eqn:Saddle}), one finds that the density
$\rho$ is $1 - z$. Therefore, the critical density is non-zero and is given by
$\rho_c = 1 - \beta$. We anticipated this result in the discussion in Section
\ref{sec:Results} in the context of the $\lambda \to 0$ transition in the BRM. 
Note that the analysis given here does not make any predictions about the
behaviour for densities below $\rho_c$. However, the results presented in
Sections \ref{sec:MFT-comp} and \ref{sec:MFT-zero}, together with the
discussion in the context of the BRM in Section \ref{sec:Results}, lead to the
conclusion that the low density phase comprises a jam in which the leading
particle hops forward with probability $\beta$ (because the site in front of it
contains passengers but no bus), and all following particles hop forward with
probability $1$, so long as the site in front contains no bus (which holds with
probability $\beta$).

\subsection{Finite Systems}

It is possible to analyse the MFM for finite systems since, using
(\ref{eqn:Pcalc}), we have
\[
	\sum_{x=0}^{L-M} p(x) = 1 = 
	\sum_{x=0}^{L-M} \frac{Z(L-x-1,M-1) q(x)}{Z(L,M)}
\]
which gives the following recurrence relation for $Z(L,M)$:
\begin{equation}
	Z(L,M) = \sum_{x=0}^{L-M} Z(L-x-1,M-1) q(x)
	\label{eqn:Zrecurrence}
\end{equation}
with the initial condition $Z(L,1) = q(L-1)$. 

In principle, one can calculate $Z(L,M)$ for any $L$ and $M$. However,
numerical precision restricts the practical maximum values of $L$ and $M$ to
$\sim 2000$. In practice for these finite system calculations, it is not
feasible to retain the dependence of $u(x)$ on $v$ and so we simply take $v =
\beta$ in the expression (\ref{eqn:HopProb}) for $u(x)$. This is a good
approximation since we are principally interested in the case where $v$ is
close to $\beta$.

\subsection{Comparison with BRM}	\label{sec:MFT-comp}

We now proceed to compare MFM results with BRM simulation data. All the
results presented for the MFM are based on a calculation of the gap size
distribution $p(x)$. It is not possible to obtain a closed analytic expression
for $p(x)$ and so one must perform the calculation numerically for a given pair
of parameters $\beta$ and $\lambda$.

\begin{figure}
	\centps{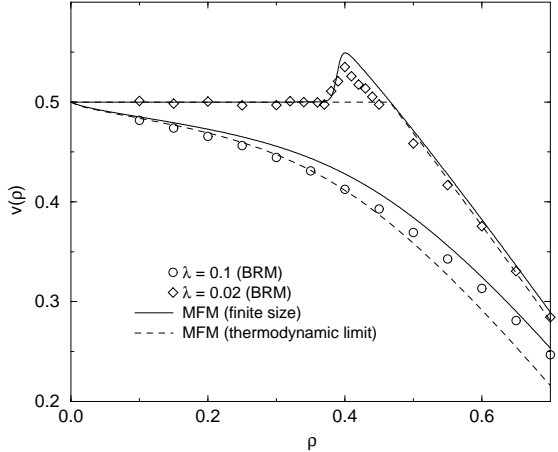}{70}
	\caption{Velocity as a function of density for the BRM (simulation) and the
	MFM (calculation) for $\lambda = 0.1$ and $\lambda = 0.02$. $\beta$ is
	$0.5$ and $L$ is $1000$. MFM data for the thermodynamic limit is also
	shown.}
	\label{fig:CompareV}
\end{figure}

Figure \ref{fig:CompareV} shows $v(\rho)$ for the BRM and the MFM; the
agreement is quite good. For $\lambda = 0.1$, we have found numerically that in
the BRM, system size is not an important factor for large enough systems and
$v(\rho)$ for $L=1000$ is essentially the same as $v(\rho)$ in the
thermodynamic limit. (The major part of the small discrepancy between the MFM
data for $L=1000$ and $L=\infty$ at high density can be attributed to the fact
that the calculation for the finite system is performed with $v$ being replaced
by $\beta$ in the expression for $u(x)$.)

Recall from Figure \ref{fig:SD-1} that in the BRM, where we had $L=10000$, for
$\lambda = 0.02$ and $\beta=0.5$, $v(\rho)$ exhibited an apparent discontinuity
in its derivative, a possible signal of a phase transition. For the MFM in the
thermodynamic limit with the same parameters, $v(\rho)$ is very similar (see
Figure \ref{fig:CompareV}) but we know that it is actually a smooth function
and that there is no transition. This is consistent with our view that there is
in fact no phase transition in the BRM for $\lambda > 0$. Figure
\ref{fig:CompareV} also shows that, for this value of $\lambda$ (unlike
$\lambda = 0.1$), there is a marked finite size effect at $L=1000$ comprising a
bump in $v(\rho)$. The bump appears for both the BRM numerics and the MFM
calculation. The reason for the presence of the bump is that the size of the
large gap in front of a jam decreases as $L$ decreases, resulting in
the ``head'' of the jam catching up with the ``tail'' (and hence an
average velocity greater than $\beta$) for densities sufficently close to
$\rho_c = 1 - \beta$.

\begin{figure}
	\centps{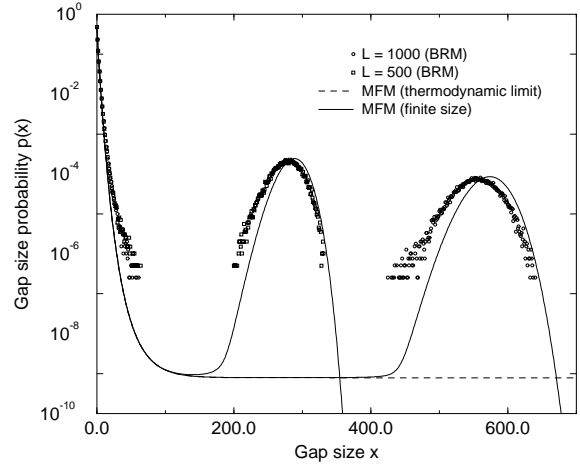}{70}
	\caption{BRM (simulation) and MFM (calculation) gap size distributions on a
	linear-log scale for small systems in the jammed regime ($\beta = 0.5$,
	$\lambda = 0.02$, $\rho = 0.2$). The MFM distribution in the thermodynamic
	limit is also shown.}
	\label{fig:GapDist-1}
\end{figure}

Figure \ref{fig:GapDist-1} compares the BRM and MFM gap size distributions for
small systems in the jammed regime. The distributions are bimodal with the
approximately Gaussian second peak corresponding to the presence of a single
large gap in the system. The agreement between simulation and MFM is again
reasonably good. The MFM distribution in the thermodynamic limit is also shown
and one can see that it decays monotonically, but very slowly, for large gap
sizes. We have found that the position of the peak in the tail of $p(x)$
increases linearly with system size as one would expect -- the peak corresponds
to the presence of an ``extensive'' gap (a gap with size $\propto L$).

\subsection{Small $\lambda$ behaviour of MFM}	\label{sec:MFT-zero}

\begin{figure}
	\centps{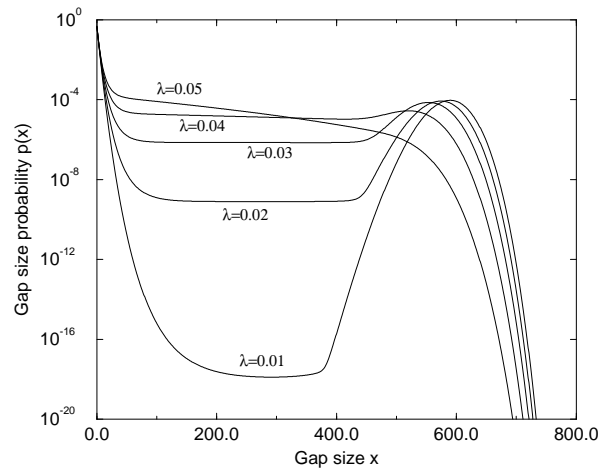}{70}
	\caption{MFM gap size distributions on a linear-log scale for a small system
	showing the effect of decreasing $\lambda$. The other parameters are $\beta
	= 0.5$, $\rho = 0.2$ and $L = 1000$.}
	\label{fig:GapDist-2}
\end{figure}

We now examine in detail the behaviour of the MFM when $\lambda$ is small (but
nonzero). Figure \ref{fig:GapDist-2} shows the effect on the MFM gap size
distribution of decreasing $\lambda$ towards zero in a finite system. For
$\lambda = 0.05$, $p(x)$ decreases monotonically from a maximum at $x=0$. As
$\lambda$ is decreased, $p(x)$ becomes bimodal, signalling the presence of
jams. The second peak becomes more pronounced as $\lambda$ is made smaller;
jamming is enhanced.

\begin{figure}
	\centps{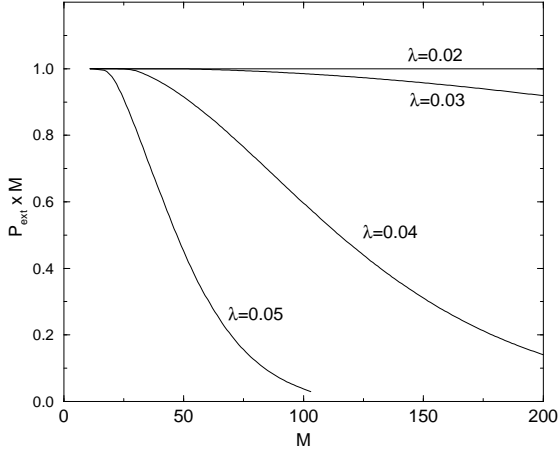}{70}
	\caption{$P_{ext} \times M$ against $M$ for the MFM showing the effect of
	decreasing $\lambda$. The other parameters are $\beta = 0.5$ and $\rho =
	0.1$.}
	\label{fig:PeakArea}
\end{figure}

In order to quantify the behaviour of $p(x)$ as $\lambda \to 0$, we define
$P_{ext}$ as the area under the peak in the tail of $p(x)$. Then $P_{ext}$ is
the probability of finding an extensive gap in the system and $P_{ext} \times
M$ is the average number of extensive gaps in the system. Figure
\ref{fig:PeakArea} shows plots of $P_{ext} \times M$ against $M$ for various
values of $\lambda$. As $\lambda$ is decreased from $0.05$, one must go to
larger systems to observe $P_{ext} \times M$ falling below $1$. This suggests
that in the limit $\lambda \to 0$, a single extensive gap survives in the
thermodynamic limit, supporting our claim that a condensation (or jamming)
transition does occur in this limit.

\begin{figure}
	\centps{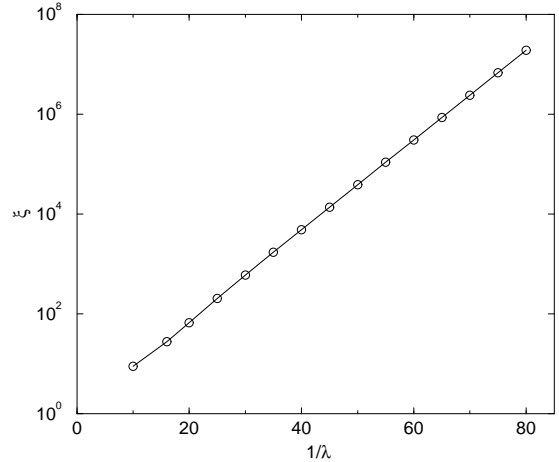}{70} 
	\caption{Linear-log plot of the typical large gap size $\xi$ against
	$1/\lambda$ for the MFM in the thermodynamic limit. The density is fixed at
	$\rho = 0.3$ and $\beta$ is $0.5$. The lines are shown to guide the eye.}
	\label{fig:Xi}
\end{figure}

We showed in Section \ref{sec:MFM-analysis} that for the MFM in the
thermodynamic limit, $p(x)$ is proportional to $(v/\beta)^x$ for large
$x$. Therefore, the typical size of the large gaps in the system is the
decay constant of $p(x)$, given by
\begin{equation}
	\xi = \left[ \ln \left( \frac{\beta}{v} \right) \right]^{-1}.
	\label{eqn:xi-def}
\end{equation}
Figure \ref{fig:Xi} shows a linear-log plot of $\xi$ against $1/\lambda$ for
$\rho = 0.3$. We see $\xi \sim \exp(a/\lambda)$ (where $a$ is a constant) for
small $\lambda$ so that the typical size of the large gaps is very large for
$\lambda$ less than about $0.02$ and has an essential singularity as $\lambda
\to 0$. Since $p(x)$ is sharply peaked at $x=0$ (see Figures 
\ref{fig:GapDist-1} and \ref{fig:GapDist-2}), we deduce that for low density 
and small $\lambda$, a MFM system comprises large clusters of buses which are
typically a distance $\xi$ apart. Clearly, this can only apply if $L
\gg \xi$; if this is not the case, then the gap size distribution must be
bimodal as seen in Figure \ref{fig:GapDist-2}.

\begin{figure}
	\centps{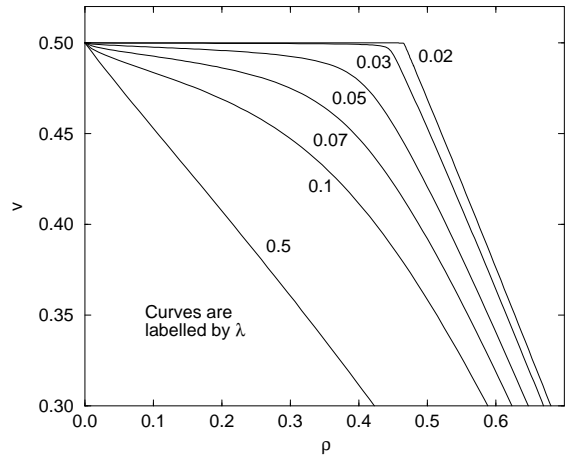}{70}
	\caption{Velocity as a function of density for the MFM in the thermodynamic
	limit showing the effect of decreasing $\lambda$. $\beta$ is $0.5$.}
	\label{fig:V-Trend}
\end{figure}

Figure \ref{fig:V-Trend} shows the effect on $v(\rho)$ of decreasing $\lambda$
towards zero in the thermodynamic limit. As $\lambda$ is decreased, a
``corner'' becomes apparent at a density slightly below $0.5$. This behaviour
is also observed in the BRM (see Figure \ref{fig:SD-1}). While we know that
there is no singularity in $v(\rho)$ for $\lambda > 0$, one can see that as
$\lambda \to 0$, $v(\rho) \to \beta$ for $\rho < 0.5$ and $v(\rho) \to 1 -
\rho$ for $\rho > 0.5$. We now examine the sharpness of the crossover from the
low density jammed regime to the high density homogeneous regime for small
$\lambda$.

\begin{figure}
	\centps{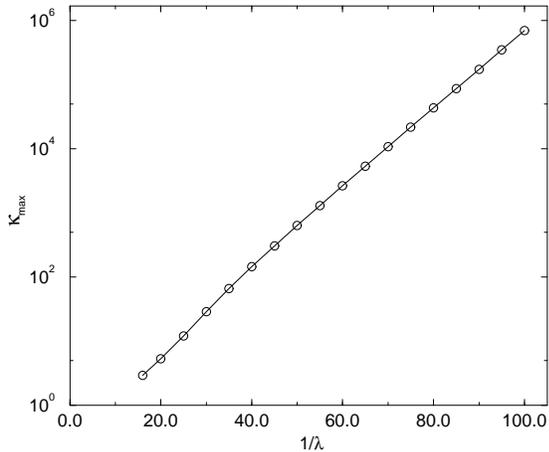}{70} 
	\caption{Linear-log plot of $\kappa_{\rm max}$ against $1/\lambda$ for the
	MFM in the thermodynamic limit. $\beta$ is $0.5$. The lines are shown to
	guide the eye.}
	\label{fig:MaxK}
\end{figure}

To estimate the sharpness of the crossover for a given set of parameters, we
calculate $\kappa_{\rm max}$, which we define to be the maximum value of
$|v''(\rho)|$. Figure \ref{fig:MaxK} shows a linear-log plot of $\kappa_{\rm
max}$ against $1/\lambda$. For $\lambda$ less than about $0.02$, we see that
$\kappa_{\rm max}$ goes as $\exp(b/\lambda)$, where $b$ is a
constant. Therefore, although a strict phase transition occurs only in the
limit $\lambda \to 0$, the crossover is exponentially sharp in $1/\lambda$ for
small $\lambda$; this may be compared with the typical large gap size $\xi$
discussed above which is also exponentially large in $1/\lambda$. For practical
purposes, the crossover may be indistinguishable from a phase transition as is
already the case for $\lambda = 0.02$ (see Figures \ref{fig:SD-1} and
\ref{fig:V-Trend}). Note that, as mentioned previously, the thermodynamic and
$\lambda \to 0$ limits do not commute: for a finite system the behaviour at
$\lambda = 0$ is trivial (there are no passengers in the system). This means
that in a large but not infinite system, the sharp crossover will, as $\lambda$
is reduced, resemble a phase transition most strongly for some nonzero
$\lambda$, $\lambda^*(L)$, and then fade away for $\lambda \ll
\lambda^*$. (Heuristically, we expect $\lambda^* L$ to be of order unity.)

\section{Coarsening}		\label{sec:Coarsening}

In this section we discuss the approach to the steady state in the jamming
regime. We have already seen in Figure \ref{fig:ST-0.02-jam} that, in finite
systems, coarsening of bus clusters (or equivalently the gaps between clusters)
occurs for small $\lambda$ and low bus density. Here, by relating the model to
a reaction-diffusion process, we argue that the typical size of the large gaps
in the system should eventually grow as $t^{1/2}$ and we provide numerical
evidence for this. We also study numerically the relaxation of the average
particle velocity $v(t)$ and find that it decays as a power law with an
exponent close to $-1$.

The gap size distribution in the jamming regime can be considered as a
superposition of the small gaps inside bus clusters and the large gaps between
these clusters. We define $r(t)$ to be the mean size of the large gaps. Recall
first from Section \ref{sec:MFM} that in a mean-field approximation we expect
the probability that a particle hops into a gap of size $r$ to be
\begin{equation}
	u(r) = \beta + ( 1 - \beta ) \exp ( -\lambda r / v )
	\label{eqn:udef-again}
\end{equation}
If $\lambda r$ is sufficiently large, the motion of a typical bus cluster
therefore becomes uncorrelated with the motion of the cluster ahead. When two
clusters come sufficiently close together, they coalesce and form a single
cluster. The larger $r(t)$ becomes, the more this coalescence is
diffusion-dominated ({\it i.e.\/} correlations in the motion of clusters are
reduced) and the more it resembles a driven $A + A \to A$ reaction-diffusion
process \cite{DbA} with bus clusters taking the place of $A$ particles. In the
driven $A + A \to A$ process\footnote{For the $A+A \to A$ process, the presence
of a preferred direction does not change the scaling from that of the undriven
system \cite{PCG}.}, the characteristic length scale (of which $r(t)$ is an
example) grows as $t^{1/2}$. We therefore expect $t^{1/2}$ growth in the BRM
when $\lambda r(t)$ is sufficiently large, {\it i.e.\/} at sufficiently late
times.

We have performed simulations on large systems ($L=5 \times 10^5$) to
investigate the approach to the steady state. In these simulations, inspection
of the complete gap size distribution shows that, after a short time, there is
a very deep minimum at a gap size of about $20$; we observe a negligible number
of gaps of this size. Therefore, in the results presented below we have defined
a large gap as one with size greater than $20$ and $r(t)$ is then the mean size
of gaps greater than $20$. In discussing the coarsening in a disordered driven
diffusive model, Krug and Ferrari \cite{KF} use the variance of the complete
gap size distribution as their measure of the typical length scale. We have
also studied this quantity and we find that its time evolution is entirely
consistent with that of $r(t)$, as one would expect in a scaling regime.

\begin{figure}
	\centps{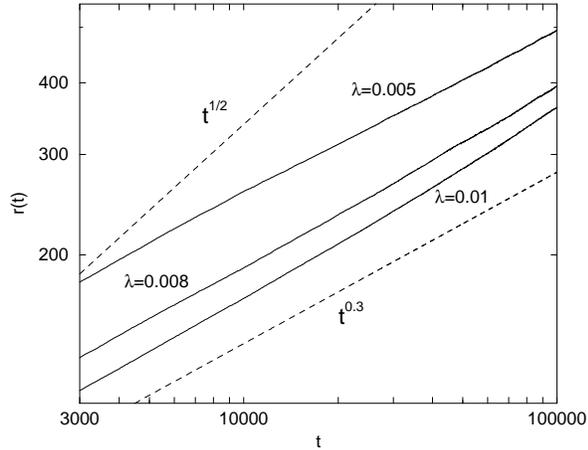}{70}
	\caption{Log-log plots of the average large gap size $r(t)$ for small 
	values of $\lambda$. Simulations had $L=500000$, $\rho = 0.2$ and $\beta =
	0.2$. Results for $\lambda = 0.01$ and $0.005$ were averaged over 15
	independent runs and for $\lambda = 0.008$, 8 runs. At $t=0$ the buses were
	positioned at random and there were no passengers. The dashed lines 
	($t^{1/2}$ and $t^{0.3}$) are shown to guide the eye.}
	\label{fig:R-scale}
\end{figure}

Figure \ref{fig:R-scale} shows log-log plots of $r(t)$ obtained from
simulation. Close inspection of the $r(t)$ curves reveals that they are
straight lines at early times indicating power law growth with exponent $\simeq
0.3$. By early times we mean the interval $3000 < t < 10000$ for the two larger
values of $\lambda$ and $10000 < t < 30000$ for $\lambda = 0.005$. However, at
later times we see different behaviour as the curves become somewhat concave,
particularly for the two larger values of $\lambda$.

Since we anticipate that $t^{1/2}$ growth will occur only after $r(t)$ has
become sufficiently large for bus clusters to be uncorrelated, we expect $r(t)$
to be of the form
\begin{equation}
	r(t) = r_0(\lambda) + A t^{1/2}
	\label{eqn:HalfFit}
\end{equation}
at late times. The constant $A$ should depend only very weakly on $\lambda$ if
the coarsening process is truly diffusion-dominated. However, $r_0$ is a
measure of how far apart bus clusters must become to be uncorrelated for a
particular value of $\lambda$, and so we expect it to vary as $\sim 1/\lambda$
from (\ref{eqn:udef-again}).

\begin{figure}
	\centps{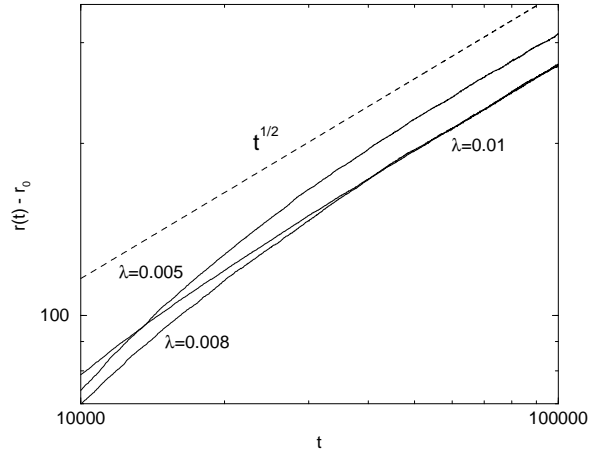}{70}
	\caption{Log-log plots of $r(t) - r_0$ for small values of $\lambda$. The
	data is the same as that presented in Figure \ref{fig:R-scale}. The dashed
	line ($t^{1/2}$) is shown to guide the eye.}
	\label{fig:HalfFit}
\end{figure}

Figure \ref{fig:HalfFit} shows log-log plots of $r(t) - r_0$ for the same data
presented in Figure \ref{fig:R-scale}. The parameter $r_0$ was estimated by
fitting the function (\ref{eqn:HalfFit}) to the data for $t > 40000$. For the
two larger values of $\lambda$ we see that the curves collapse at late times as
expected. The behaviour at these late times is certainly consistent with
$t^{1/2}$ growth. However, the results for $\lambda = 0.005$ do not fit this
picture. We believe that this is because the $t^{1/2}$ growth regime has not
yet been reached for this value of $\lambda$. It appears that as $\lambda$
becomes smaller, the regime characterised by approximate $t^{0.3}$ growth
increases in duration. Our results suggest that the BRM may exhibit a crossover
from power law growth of $r(t)$ with exponent $\simeq 0.3$ to growth with
exponent $1/2$. We remark that the MFM exhibits similar, but not identical,
coarsening behaviour.

\begin{figure}
	\centps{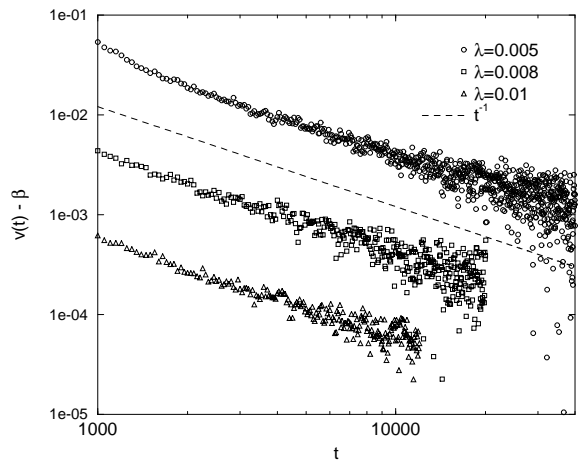}{70}
	\caption{Log-log plots of $v(t) - \beta$ for small values of $\lambda$.
	Simulations are the same as those for the data shown in Figures
	\ref{fig:R-scale} and \ref{fig:HalfFit}. The dashed line ($t^{-1}$) is 
	shown to guide the eye.	The data has been rescaled vertically for clarity.}
	\label{fig:V-scale}
\end{figure}

We now discuss the behaviour of the average velocity $v(t)$ as the steady state
is approached. We believe that the true steady-state velocity in the jammed
regime of the BRM is very close to, but slightly smaller than, $\beta$. Figure
\ref{fig:V-scale} shows the decay of $v(t)$ towards its steady-state value. 
The data, although noisy, shows clearly that $v(t)$ relaxes towards $\beta =
0.2$ as a power law with an exponent close to $-1$. The data for later times,
although very noisy and not shown, is consistent with $t^{-1}$. We have no
evidence for a crossover in the relaxation of $v(t)$ (in contrast to $r(t)$)
but, owing to the quality of our data, neither can we rule out this
possibility.

It is interesting to note that we observe coarsening in a system which does not
strictly phase separate. We have already argued that for small $\lambda$ the
thermodynamic singularities in the limit $\lambda \to 0$ are replaced by
behaviour that is smooth but has crossovers that are exponentially sharp in
$1/\lambda$. Apparently the $\lambda \to 0$ phase transition likewise shows up
in the dynamics where the system appears to phase separate, but only up to a
finite (but very long) time so that a truly inhomogeneous state is never
reached. Our study of the MFM for small $\lambda$ in Section \ref{sec:MFT-zero}
suggests, however, that the final ``homogeneous'' state comprises large
clusters of buses separated by gaps having a typical size of order
$\exp(1/\lambda)$; hence $t^{1/2}$ growth of the typical large gap size will
occur up to some time exponentially large in $1/\lambda$ (for a sufficiently
large system). Perhaps not surprisingly, this timescale is of the same order of
magnitude as our estimate of the timescale on which individual clusters are
stable, derived within the two-particle approximation in Section
\ref{sec:TwoBody}.

\section{Dual Model: Clogging}	\label{sec:Holes}

The above completes our study of the BRM. However, as mentioned in
Section \ref{sec:Introduction}, our motivation for studying the model is mainly
connected with its interesting generic behaviour (a driven system with one
conserved and one non-conserved variable) rather than its applicability to
public transport. In this section we present an alternative interpretation of
the model which further reveals its generic behaviour.

One can interpret the BRM in a different way by noting that each
time a bus hops to the right, the site (hole) that the bus hops into moves to
the left. By considering the ``holes'' (henceforth we call them {\em
particles}) to be the moving entities in the model and the ``buses'' to be
empty sites, one defines a dual model.

A feature of the dual model is that the non-conserved variable can now be
thought of as being ``attached'' to the moving particles rather than fixed
sites as in the original interpretation. The non-conserved variable is $\mu_i$,
the speed of particle $i$, which can either be fast ($\mu_i = 1$) or slow
($\mu_i = \beta$). If the dynamics of this dual model is to be exactly the same
as that of the BRM, then a particle should attempt to hop to the left when the
site to its left is updated. However, a more natural dynamics is to only choose
particles for update. The full update rules in the latter case are:
\begin{enumerate}
	\item Pick a particle $i$ at random. 
	\item If $\mu_i = 1$ then $\mu_i \to \beta$ with probability $\lambda$.
	\item If there is no particle on the site to the left of particle $i$, then
	it hops to the left with probability $\mu_i$.
	\item If particle $i$ hops, then $\mu_i \to 1$.
\end{enumerate}
This model describes a system of particles each of which can exist in two
states of mobility. A particle has probability $\lambda$ of switching to the
less mobile state each time-step for which it remains stationary. When it
finally does move, it is restored to the more mobile state.

While the dual model is not identical to the BRM, we have checked that the
numerical behaviour is indeed very similar. The density of particles is $1 -
\rho \equiv \rho'$ and we note that a gap in the BRM is equivalent to a cluster
of adjacent particles in the dual model. Thus, the mean-field theories for the
two models are equivalent, so long as the hopping rate into a gap of size $x$
in the MFM is interpreted as the hopping rate of a particle leaving the left
edge of a cluster of size $x$ in the dual mean-field theory. Jamming becomes a
{\em high density} phenomenon here, characterised by the presence of large
clusters of particles. This restores to the word ``jamming'' a meaning closer
to that used in everyday life.

We define the average speed of particles as $v'$. Since in the BRM
the magnitude of the bus current must be equal to the magnitude of the hole
current, we have
\begin{equation}
	v'  = \left( \frac{\rho}{1-\rho} \right) v(\rho) = 
		  \left( \frac{1-\rho'}{\rho'} \right) v(1-\rho').
\end{equation}

\begin{figure}
	\centps{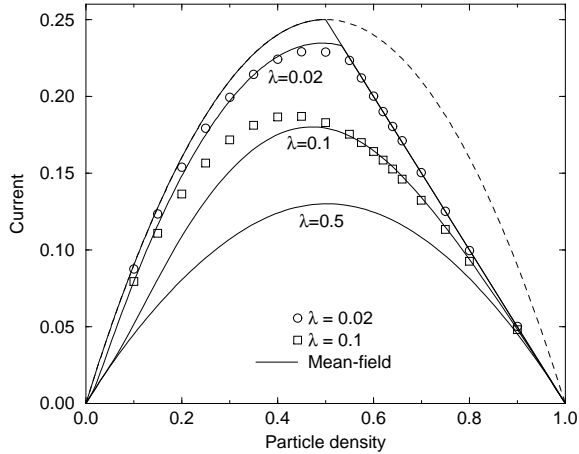}{70}
	\caption{Fundamental diagram for moving holes model for different values of
	$\lambda$. For all simulation data, $L=10000$ and MFM results are in the
	thermodynamic limit. The uppermost solid curve is the MFM result in the
	thermodynamic limit followed by the limit $\lambda \to 0$. The
	dashed curve is the exact result when $\lambda$ is set equal to zero before
	the thermodynamic limit is taken. The latter two curves are identical for
	$\rho < 0.5$.}
	\label{fig:Fundamental}
\end{figure}

The dual model can be thought of as describing stop-start traffic flow with the
particles representing cars. The longer a car is at rest (usually because a car
in front is blocking it), the more likely it is that the driver will be slow to
react when it is possible to move again. Now $\lambda$ is the parameter
determining the strength of the ``slow-to-start'' behaviour of cars. This is
similar to several ``slow-to-start'' cellular automaton traffic models studied
recently \cite{SS}. In the study of traffic flow, the so-called ``fundamental
diagram'' (the current $v'\rho'$ as a function of density $\rho'$) is commonly
examined. Figure \ref{fig:Fundamental} shows the fundamental diagram for the
dual model (simulation and MFM) for different values of $\lambda$. One can see
that for high densities and $\lambda$ very small, the current is significantly
less than that for $\lambda = 0$. It follows from the behaviour as $\lambda
\to 0$ that in the model, an infinitesimal probability for cars to become
slow-to-start results in a macroscopically large decrease in the
current at high density.

A different interpretation of the dual model provides a simple picture of
another familiar kind of jamming which might be called ``clogging''. Consider
particles suspended in a fluid which is pumped along a narrow pipe. Imagine
that the particles, which are comparable in size to the pipe width (so they
cannot pass each other), have some tendency to stick to the walls of the pipe,
but that this process takes time to occur and therefore can only happen if a
particle remains stationary with respect to the pipe wall for a significant
time. A stuck particle will remain stuck if there is another such particle in
front of it. If not, the stuck particle will detach from the wall and move on,
but only after a delay. This offers an interpretation of the dual model in
which the attachment rate of a stationary particle is $\lambda$ and the rate at
which a stuck particle will detach and move on is $\beta$. (The parameter
$\alpha$ is set by the flow rate of the fluid.) In the limit $\lambda\to 0$, a
phase transition then arises from a homogeneous phase in which the particles
move quickly, to a jammed phase of inhomogeneous, slow-moving particles. This
transition occurs when the particle density is raised above $1-\rho_c$, where
$\rho_c$ was defined earlier for the BRM.

More generally, such a model could be taken as a highly simplified discrete
description of any fluid that has a tendency to clog. There are many fluids
which will solidify when at rest but remain in a fluid state if kept moving
rapidly enough -- examples include colloid/polymer mixtures \cite{PPHP},
clay gels (used as drilling muds in the oil industry) and, in some
circumstances, blood. A similar description might apply to
particulate suspensions (say in a horizontal pipe) which, if not kept moving,
will settle under gravity into a relatively immobile deposit.

\section{Relevance to Real Buses}		\label{sec:RealBuses}

While the BRM is not designed to model a real bus route with any great
accuracy, it is worth commenting on its possible relevance to the bus route
problem. To do this, we must relate our model parameters to those of a real (in
our case, Scottish) bus route. $L$ is the number of stops on the circular route
and the bus density $\rho$ is the number of buses per stop. We expect $\rho$ to
be quite low (perhaps $0.1$), and hence to be in the regime where jamming could
potentially occur. The parameter $\alpha$, which we take to be $1$, is
inversely proportional to the average time a bus takes to travel from one stop
to the next (with no stopping for passengers). The parameters $\beta$ and
$\lambda$ are defined relative to $\alpha$. The amount by which buses are
delayed owing to their having to pick up passengers is reflected in
$\beta$. Our choice of $\beta = 0.5$ is, we believe, a reasonable figure for a
city bus route -- a bus picking up passengers at every stop will progress about
half as quickly as one which has no passengers to collect. We interpret
$\lambda$ as the probability that one or more passengers arrive at a typical
bus-stop in the time it takes a bus to go from one stop to another (roughly a
minute, say). Of course, there is no such thing as a typical bus-stop on a city
bus route but we believe that $\lambda$ should often lie between $0.1$ and $1$,
depending on location, time of day and other factors.

\begin{figure}
	\centps{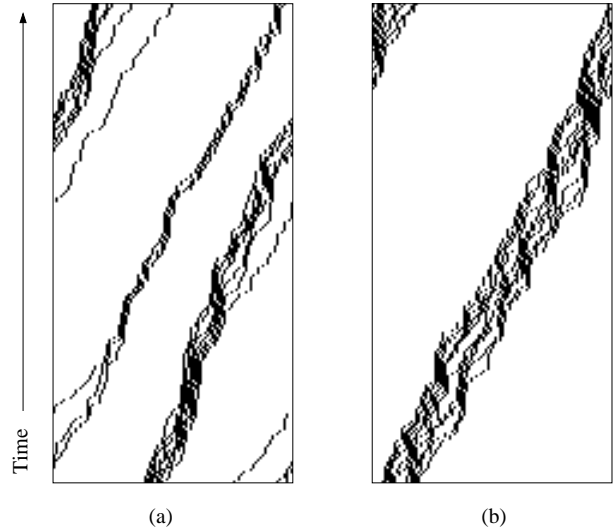}{70} 
	\caption{Space-time plots of bus positions for $\rho = 0.2$, $\beta = 0.5$
	and $L=100$. In (a), $\lambda = 0.1$ and in (b), $\lambda = 0.02$. There is
	1 time-step between each snapshot on the time axis. The system is in the
	steady state.}
	\label{fig:ST-realbus}
\end{figure}

In our examination of the BRM up to this point, we have primarily studied
the limit of large system size $L$. For real bus routes, however, we expect
$L$, the number of bus-stops, to be in the region of $50$ to $100$. Figure
\ref{fig:ST-realbus} (a) shows the positions of $10$ buses in a system of $100$
bus-stops for two hundred timesteps in the steady state, with $\lambda =
0.1$. Transient clusters of buses are clearly visible in the system, a scenario
familiar to regular users of the bus route. A similar space-time plot is shown
for $\lambda = 0.02$ in Figure \ref{fig:ST-realbus} (b). As was the case for
larger systems, we see a single large cluster of buses which is relatively
stable. Our model suggests, therefore, that such catastrophic clusters of buses
will not form if the arrival rate of passengers is sufficiently large, but that
at some intermediate arrival rates, a given bus is quite likely to be part of a
small cluster. Of course, as the arrival rate is increased still further, the
BRM predicts an increasingly disordered bus route with little clustering.

Finally, we wish to comment on the implications of a ``hail-and-ride'' bus
system, where there are no fixed bus-stops and passengers may hail a bus at any
point on the route. In the BRM, this corresponds to a system with a large
number of stops (large $L$) and very small values of both $\lambda$ and
$\beta$. Hence, one expects the buses to cluster very strongly, much as in
Figure \ref{fig:ST-0.1-0.002} (with a much lower bus density, however).
Compounding this problem, with $\beta$ very small the leading bus in a cluster
moves extremely slowly. Thus, the use of a ``hail-and-ride'' system, while on
the surface more convenient for passengers, would have dire consequences for
the efficiency of the bus service. However, it is worth pointing out that on
routes used by very few passengers, $\lambda$ might be so small that one would
observe essentially $\lambda = 0$ behaviour, {\it i.e.\/} a homogeneous system
with buses being scarcely delayed. This is because the probability that a site
has passengers on it cannot be greater than $\lambda L$; if $\lambda L
\ll 1$, no bus is significantly delayed by having to pick up passengers.

\section{Discussion}		\label{sec:Discussion}

In this work we have provided strong evidence that the BRM undergoes a jamming
transition as a function of the density $\rho$ of buses in the limit where
$\lambda$, the rate of arrival of passengers, tends to zero. This provides an
example of a homogeneous system with local, stochastic dynamics that exhibits a
jamming transition in one dimension. Although it remains an open problem to
find an exact solution of the model, we have shown that the steady state of a
mean-field approximation is solvable. This approximation captures the essence
of how the ordering into jams occurs -- the density of passengers at a site
provides information about the time at which a bus last visited the
bus-stop. From the elapsed time it is inferred how far away the next bus is
along the route. Thus the passengers mediate an effective long range
interaction between the buses. The nature of the mean-field approximation is to
replace this ``induced'' interaction (which is subject to stochastic variation)
with a deterministic one.

The transition also has strong analogies with the jamming transition induced by
a single defect particle or disorder
\cite{JL,Schutz,DJLS,Mallick,Derrida,TZ,Evans,KF}. In that case it has been 
shown that the transition is reminiscent of Bose condensation
\cite{Evans}. In the present model an interesting point is that there is no
defect bus present; all buses are equal. Instead, the system spontaneously
selects a bus behind which a jam forms and the waiting passengers condense in
front of this bus. As mentioned in Section \ref{sec:Introduction}, it would
make very little difference if overtaking of buses were allowed in the model,
since its effect would be merely to interchange the leading two buses of a
jam. The new leading bus would then proceed as slowly as its predecessor. This
makes the physics different from the defect mediated case where unhindered
overtaking would prevent a jam ever arising. We have shown, at least in the two
body approximation of Section \ref{sec:TwoBody}, that the time for the lead bus
to escape from its jam diverges strongly with system size. This may be compared
with the ``flip time'' in another model exhibiting spontaneous symmetry
breaking \cite{GLEMSS}.

The bus route model has an interesting dual model, described in Section
\ref{sec:Holes}. This considers the spaces between buses to be moving
entities. If the moving entities are interpreted as cars, the dual model is a
particular type of slow-to-start traffic model \cite{SS}. If the moving
entities are particles suspended in a fluid, the dual model is a model of
``clogging" in the transport of sticky particles, or a gelling fluid, down a
pipe. In either interpretation, there is (in the limit of small $\lambda$) a
phase transition between a homogeneous and an inhomogeneous phase; however, the
inhomogeneous (jammed) phase now arises at high density. This contrasts to the
BRM itself and related models \cite{JL,Schutz,DJLS,Mallick,Derrida,TZ,Evans,KF}
of defect-mediated jamming, for which jams arise when the vehicle density is
too low.

The dual model is, in our view, interesting because it describes the jamming of
particles whose mobility depends on an internal dynamical degree of
freedom. This is represented by the nonconserved variable $\mu$. In the dual
model itself, the nonconserved variable keeps an internal record of how long it
is since the particle last moved. More generally however, other models of this
type could entail a nonconserved internal variable representing, say, the
orientation of rodlike particles. (Such particles would be much more likely to
jam in some orientations than others.)  We have shown that the existence of an
additional, nonconserved degree of freedom is, at least in a specified limit
($\lambda \to 0$), enough to cause a symmetry-breaking phase transition to a
jammed state, in a homogeneous one dimensional driven system.

We now comment on our results for nonzero $\lambda$. Although we argue that a
transition only occurs as $\lambda \to 0$, we have shown that a strong vestige
of the transition remains for small values of $\lambda$. In fact, in the
mean-field theory of the BRM, the crossover between the two regimes becomes
exponentially sharp in $1/\lambda$. Therefore, in practice the crossover
becomes very difficult to distinguish from a strict thermodynamic transition
when $1/\lambda$ is of order $100$ or more. The $\lambda \to 0$ transition also
strongly influences the {\em dynamical} behaviour of the system when $\lambda$
is small but non-zero: we observed coarsening behaviour (presumably transient)
as would usually be associated with systems that do strictly phase
separate. All this highlights the fact that that care is required to
unambiguously identify phase transitions, as opposed to crossover
phenomena. Indeed, it is possible that closely related crossover phenomena
occur in cellular automata models of traffic \cite{ESSS,SK}.

\begin{figure}
	\centps{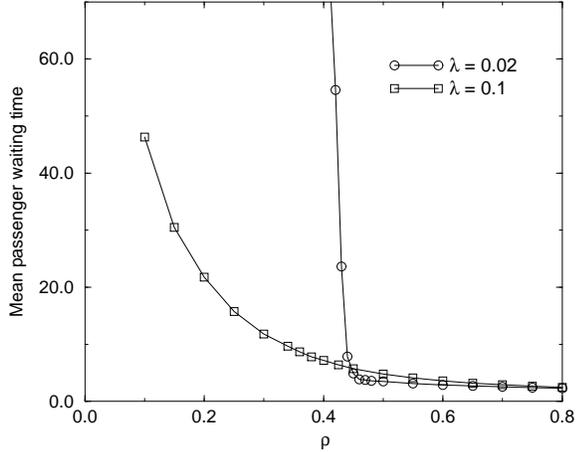}{70}
	 \caption{Plot of the mean passenger waiting time as a function
	 of bus density for two different values of $\lambda$. The simulation
	 parameters are $L=10000$ and $\beta = 0.5$.}
	\label{fig:WaitingTime}
\end{figure}

Finally, we show in Figure \ref{fig:WaitingTime} the average time that a
passenger is required to wait for a bus. The system size is $10000$, unphysical
in terms of real bus routes but representative of the thermodynamic limit. For
$\lambda = 0.1$, we see that, as one would expect, the average waiting time
increases smoothly as the density of buses is reduced. However, for $\lambda =
0.02$, it increases very sharply at an intermediate value of the density, and
for low densities, the bus service becomes highly inefficient. This figure
serves as yet another cruel reminder of the vagaries of the bus route to those
whose lives are unfettered by the ownership of a motor car.

\section{Acknowledgements}

OJO is supported by a University of Edinburgh Postgraduate Research
Studentship. MRE is a Royal Society University Research Fellow.

%###############################################################################
% Appendix
%###############################################################################

\appendix

\section{Steady State Solution of Mean Field Model}	\label{sec:ExactSoln}

The MFM comprises $M$ particles hopping on a $1d$ ring of $L$ lattice
sites. The probability of hopping into a gap of size $x$ is $u(x)$ which must
obey $u(0) = 0$ (hard-core exclusion). The update is random sequential so that
the probability of a given particle being updated at each step is
$\frac{1}{M}$.

A configuration ${\cal C}$ is completely specified by a sequence of gap sizes
$\{x_i\}$ where $x_i$ is the size of the gap in front of the $i^{\rm th}$
particle.

In the steady state, the master equation for the system is
\begin{equation}
	\sum_{{\cal C}' \not= {\cal C}} W({\cal C} \to {\cal C}') 
	P({\cal C}) =
	\sum_{{\cal C}' \not= {\cal C}} W({\cal C}' \to {\cal C}) 
	P({\cal C'})
	\label{eqn:Master1}
\end{equation}
where $P({\cal C})$ is the steady state probability of configuration
${\cal C}$ and $W({\cal C} \to {\cal C}')$ is the probability of an
elementary transition from ${\cal C}$ to ${\cal C}'$.

First, consider the RHS of (\ref{eqn:Master1}). Define ${\cal C} = \{ x_1,
\ldots, x_i, \ldots, x_M \}$ and ${\cal C}_j = \{ x_1, \ldots, x_{j-1}-1,
x_j + 1, \ldots, x_M \}$. Then the only configurations from which
${\cal C}$ can be obtained by an elementary transition are the
${\cal C}_j$, with the constraint that $x_{j-1} > 0$. We therefore
have
\begin{equation}
	W({\cal C}_j \to {\cal C}) = \frac{1}{M} u(x_j + 1)
\end{equation}
and the RHS of (\ref{eqn:Master1}) becomes
\begin{equation}
	\frac{1}{M} \sum_{j} u(x_j + 1) P({\cal C}_j) \theta(x_{j-1})
	\label{eqn:RHS}
\end{equation}
where $\theta(x)$ is the usual Heaviside (step) function.

Now consider the LHS of (\ref{eqn:Master1}). Define ${\cal C}_k = \{ x_1,
\ldots, x_{k-1} + 1, x_k - 1, \ldots, x_M \}$. The ${\cal C}_k$ are the only
configurations which can be obtained from ${\cal C}$ by an elementary
transition, with the constraint that $x_k > 0$. Therefore we have
\begin{equation}
	W({\cal C} \to {\cal C}_k) = \frac{1}{M} u(x_k)
\end{equation}
and the LHS of (\ref{eqn:Master1}) becomes
\begin{equation}
	\frac{1}{M} \sum_{k} u(x_k) P({\cal C}) \theta(x_k).
	\label{eqn:LHS}
\end{equation}

Equating (\ref{eqn:LHS}) and (\ref{eqn:RHS}) gives
\begin{equation}
	\sum_{k} u(x_k) P({\cal C}) \theta(x_k)=
	\sum_{j} u(x_j + 1) P({\cal C}_j) \theta(x_{j-1}).
	\label{eqn:Master2}
\end{equation}
Since periodic boundary conditions apply, the indices on the LHS and RHS of
(\ref{eqn:Master2}) can be matched up to give
\begin{equation}
	\sum_{j} u(x_{j-1}) P({\cal C}) \theta(x_{j-1}) =
	\sum_{j} u(x_j + 1) P({\cal C}_j) \theta(x_{j-1}).
	\label{eqn:Master3}
\end{equation}

To solve this equation, we assume a product form for $P$. We write
\begin{equation}
	P(\{x_i\}) = \frac{1}{Z(L,M)} q(x_1) \ldots q(x_i) \ldots q(x_M)
	\label{eqn:Ansatz}
\end{equation}
where $Z(L,M)$ is a normalisation. Substituting this product form into
(\ref{eqn:Master3}) gives
\begin{equation}
	\sum_{j} \left( \prod_{i} q(x_i) \right) \theta(x_{j-1}) A_j = 0
\end{equation}
where
\begin{equation}
		A_j = u(x_{j-1}) - 
			\frac{u(x_j+1) q(x_{j-1}-1) q(x_j+1)}{q(x_{j-1}) q(x_j)}.
\end{equation}
To solve this for $q(x_j)$, make the ansatz that
\begin{equation}
	u(x_{j-1}) \frac{q(x_{j-1})}{q(x_{j-1}-1)} = 
	u(x_j+1) \frac{q(x_j+1)}{q(x_j)}
\end{equation}
for $x_{j-1} > 0$. This has the solution
\begin{equation}
	q(x) = \prod_{y=1}^x \frac{1}{u(y)}
\end{equation}
which, together with (\ref{eqn:Ansatz}), gives $P(\{x_j\})$ as required.

\end{document}